%% file: Overview.tex
\documentclass[pdflatex,sn-mathphys-ay]{sn-jnl}

\usepackage{graphicx}%
\usepackage{multirow}%
\usepackage{amsmath,amssymb,amsfonts}%
\usepackage{amsthm}%
\usepackage{mathrsfs}%
\usepackage[title]{appendix}%
\usepackage{xcolor}%
\usepackage{textcomp}%
\usepackage{manyfoot}%
\usepackage{booktabs}%
\usepackage{algorithm}%
\usepackage{algorithmicx}%
\usepackage{algpseudocode}%
\usepackage{listings}%
\usepackage{tabularx,booktabs}
\usepackage{adjustbox}
\usepackage{longtable}
\usepackage{CJKutf8}

\theoremstyle{thmstyleone}%
\theoremstyle{thmstyletwo}%

\theoremstyle{thmstylethree}%

\begin{document}

\title[Article Title]{An Overview of Exocomets}


\author*[1]{\fnm{Daniela} \sur{Iglesias}}\email{D.P.Iglesias@leeds.ac.uk}

\author[2]{\fnm{Isabel} \sur{Rebollido}}

\author[3,4]{\fnm{Azib} \sur{Norazman}}

\author[5]{\fnm{Colin} \sur{Snodgrass}}

\author[6,7]{\fnm{Darryl Z.} \sur{Seligman}}

\author[8]{\fnm{Siyi} \sur{Xu} \begin{CJK*}{UTF8}{gbsn}(许\CJKfamily{bsmi}偲\CJKfamily{gbsn}艺\end{CJK*})}

\author[9]{H.~\fnm{Jens} \sur{Hoeijmakers}}

\author[10]{\fnm{Matthew} \sur{Kenworthy}}

\author[11]{\fnm{Alain} \sur{Lecavelier des Etangs}}

\author[12]{\fnm{Michele} \sur{Bannister}}

\author[13]{\fnm{Bin} \sur{Yang}}


\affil*[1]{\orgdiv{School of Physics and Astronomy}, \orgname{University of Leeds}, \orgaddress{\street{Sir William Henry Bragg Building}, \city{Leeds}, \postcode{LS2 9JT}, \country{UK}}}

\affil[2]{\orgdiv{European Space Astronomy Centre (ESAC)}, \orgname{European Space Agency (ESA)}, \orgaddress{\street{Camino Bajo del Castillo s/n}, \city{Villanueva de la Ca\~{n}ada}, \postcode{28692}, \state{Madrid}, \country{Spain}}}

\affil[3]{\orgdiv{Department of Physics}, \orgname{University of Warwick}, \orgaddress{\street{Gibbet Hill Road}, \city{Coventry}, \postcode{CV4 7AL}, \country{UK}}}

\affil[4]{\orgdiv{Centre for Exoplanets and Habitability},\orgname{University of Warwick}, \orgaddress{\street{Gibbet Hill Road}, \city{Coventry}, \postcode{CV4 7AL},\country{UK}}}

\affil[5]{\orgdiv{Institute for Astronomy}, \orgname{University of Edinburgh}, \orgaddress{\street{Royal Observatory}, \city{Edinburgh}, \postcode{EH9 3HJ}, \country{United Kingdom}}}

\affil[6]{\orgdiv{Department of Physics and Astronomy}, \orgname{Michigan State University}, \orgaddress{ \city{East Lansing}, \postcode{48824}, \state{MI}, \country{USA}}}

\affil[7]{\orgdiv{NSF Astronomy and Astrophysics Postdoctoral Fellow}}

\affil[8]{\orgname{Gemini Observatory/NOIRLab}, \orgaddress{\street{950 N Cherry Ave}, \city{Tucson}, \postcode{85719}, \state{AZ}, \country{USA}}}

\affil[9]{\orgdiv{Division of Astrophysics, Department of Physics}, \orgname{Lund University}, \orgaddress{\street{Professorsgatan 1B}, \city{Lund}, \postcode{221 00}, \country{Sweden}}}

\affil[10]{\orgdiv{Leiden Observatory}, \orgname{Leiden University}, \orgaddress{\street{Einsteinweg 55}, \city{Leiden}, \postcode{NL-2333}, \country{The Netherlands}}}

\affil[11]{\orgdiv{Institut d'astrophysique de Paris}, \orgname{CNRS-Sorbonne Universit\'e}, \orgaddress{\street{98bis boulevard Arago}, \city{Paris}, \postcode{75014}, \country{France}}}

\affil[12]{\orgdiv{School of Physical and Chemical Sciences--Te Kura Mat\=u}, \orgname{University of Canterbury}, \orgaddress{\street{Private Bag 4800}, \city{Christchurch}, \postcode{8140}, \country{New Zealand}}}

\affil[13]{\orgdiv{Instituto de Estudios Astrofísicos, Facultad de Ingeniería y Ciencias}, \orgname{Universidad Diego Portales}, \orgaddress{\street{Av. Ej\'ercito Libertador 441}, \city{Santiago},  \postcode{8370191}, \country{Chile}}}




\abstract{We give a general overview of what the scientific community refers to as ``exocomets". The general definition of exocomets, as presented in this work, is discussed and compared with Solar System comets and interstellar objects, addressing their detection around main-sequence stars as well as orbiting white dwarfs. We introduce the different types of exocomet observations, highlighting the difference between exocometary `bodies' and exocometary `material'. We provide a census of all exocometary system candidates detected so far, both via spectroscopy and photometry, including detections around white dwarfs.}






\maketitle

\section{Introduction}
\label{sec:intro}

Small bodies in the Solar System are considered fossils of the planet formation process, holding key information about the chemistry and dynamics of its early stages. 
They are compact reservoirs of solid material that can be transported throughout planetary systems, particularly volatiles, as well as the delivery of ices to the inner system \citep{Obrien2018}. 
Their role in collisional processing is crucial to understanding both the age and evolution of surfaces throughout the Solar System, and the history of life on Earth \citep{Alvarez1984}.

Small bodies in other stellar systems should therefore provide similarly useful information for planet formation and evolution throughout the Galaxy. 
The astronomical community has detected extrasolar small body material in the form of exocomets \citep{Kiefer2014} --- which form the main focus of our discussion ---, exocometary residuals (such as gas and dust leftover from former exocomets; \citealt{Dent_2014}), interstellar objects \citep{Fitzsimmons2023}, and interplanetary dust \citep{Sterken2019}.

Nowadays, the word `exoplanet' is widely accepted in the astronomical community and familiar to the general public. 
With over 6000\footnote{\url{https://science.nasa.gov/exoplanets/} 6028 confirmed exoplanets, accessed on the 17$^{\rm th}$ of October, 2025.} confirmed exoplanets as of October 2025 (and counting), their number has increased dramatically during the decade 2010-2020. 
However, their discovery is fairly recent in the history of astronomical research, with the first discovery of an exoplanet around a solar-type star occurring in 1995 (51 Pegasi b; \citealt{MayorQueloz1995}). 
Often overlooked is that the discovery of small bodies orbiting other stars predated that of extrasolar planets. 
In 1987, \citeauthor{Ferlet_1987} reported variations in the Ca\,{\sc ii} K line of $\beta$~Pictoris and proposed to explain them by cometary-like objects transiting the star. Many studies followed, confirming the observed variations in the Ca\,{\sc ii} K (and H) line of $\beta$ Pic, also observed in other ionised lines such as Mg\,{\sc ii}, Fe\,{\sc ii}, and Al\,{\sc iii} (e.g. \citealt{Lagrange1987}, \citealt{Lagrange-Henri1988}, \citealt{Beust1991}, \citealt{Deleuil1993}, \citealt{Vidal-Madjar_1994}, \citealt{Lagrange1996}, \citealt{Kiefer2014}, \citealt{Tobin2019}). 
The term `FEBs' (Falling Evaporating Bodies) was used in some works since \cite{Beust1990} to refer to these infalling planetesimals (which in reality are sublimating rather than evaporating). 
Today, these are more frequently termed `exocomets', analogous to `exoplanets': `comets' around other stars. 
However, the exact meaning of the term `exocomet' is not yet well defined, and in Section \ref{sec:def} we address its definition.

Exoplanets are expected to be as common as stars; population studies show that the average number of planets per star is close to one \citep{Mann2017,Livingston2018,Kruse2019,Schulte2024}. Small bodies, including exocometary nuclei, should therefore also be common given that they are a natural by-product of the planetary formation process. However, our understanding of exocomets is limited to a handful of detections.  Most of what we know is based on observations and modelling of a single system: $\beta$~Pictoris. This system is an A6V-type star hosting two planets; b$\sim$12 M$_{\mathrm {Jup}}$ and c$\sim$10 M$_{\mathrm {Jup}}$ (\citealt{Lagrange2009,Lagrange2019}) and surrounded by an edge-on debris disc \citep{SmithTerrile1984}.  $\beta$~Pictoris is considered to be the canonical exocometary host system and has provided thousands of exocomet detections at different wavelengths. This system is quite young, with an estimated age of 23$\pm$3 Myr \citep{MamajekBell2014}, and bright; V$_{\rm mag}$=3.86 \citep{Ducati2002}, and is located at only 19.63 pc from our Solar System \citep{GaiaEDR3_2020}. Spectroscopic observations show that exocometary transits occur almost daily in $\beta$~Pictoris and these transits have been confirmed with photometric observations \citep{Kiefer2014, Tobin2019, Zieba2019}. This system has also yielded the largest number of chemical species detected in exocomets. The distinctiveness of this system may arise from the rarity of bright and young edge-on systems. For a more comprehensive review of $\beta$~Pictoris we refer the reader to Lu et al. (submitted). We provide this brief overview because $\beta$~Pictoris will be frequently mentioned as the reference case in these series due to its uniqueness. 

The study of exocomets has also expanded the search for circumstellar gas around main-sequence stars as, in some cases, this gas is believed to originate from exocomets \citep[e.g.][]{Matra2017a, Kral2019}.
Most protoplanetary disc material should be depleted at the main-sequence stage, and dust would be generated by collisional cascades \citep{Wyatt2008}. In addition to the early detections of CO through UV spectroscopy \citep{Vidal-Madjar_1994, Jolly1998, Roberge2000}, the advent of the Atacama Large Millimeter Array (ALMA) and other interferometric facilities provide evidence that there are non-negligible gas reservoirs at the main-sequence stage.  It is currently unclear whether this is primordial gas left over from the protoplanetary disc stage, or secondary gas released by collisions and/or sublimating bodies \citep[e.g.][]{Moor2017,Matra2017b,Marino_2020, Nakatani2021}. There are convincing arguments and inconsistencies for both origin theories. However, in the case the gas is proven to be secondary, then observations of this gas would provide an indirect method to study the composition of icy and rocky bodies in planetary systems. Moreover, other studies have expanded to post-main sequence stages, where the compositions of rocky bodies are investigated through the pollution on white dwarfs \citep{Rogers2024a} and their circumstellar dust and gas \citep{Dennihy2020b}. 

This introductory work sets the scene for a series of publications about exocomets, their origin, evolution, and relevance in the study of planetary systems. A list of these manuscripts and a brief description is provided at the end of this work.


\section{Comets in extra-solar systems}\label{sec:def}


In this section, we define what we mean by the term `exocomet' for the purposes of this work and others that are part of this collection. To begin with, it is worth noting that the term `comet' does not have a rigid definition within studies of our own Solar System, and consequently the word `exocomet' has been used quite broadly in the literature. Therefore, we start this section by discussing the definition of Solar System comets, followed by interstellar objects, to finally address the definition of exocomets.



\subsection{Solar System Comets}
\label{Solar System comets}

Recent studies of Solar System minor bodies increasingly show that there is a continuum from rocky asteroids through to icy comets, and populations do not simply divide into one category or the other. 
Active asteroids and Main Belt comets are bodies on $\sim1\leq a \leq 5$~au orbits that show comet-like activity, normally in the form of a dust tail. 
Initial disagreement about which of these terms to use has developed into a general consensus that `active asteroids' is the more general term, meaning any body in an asteroid-like orbit that shows evidence for mass loss, while `Main Belt comets' are the subset of these that have orbits within the main asteroid belt between Mars and Jupiter, and activity suspected to be driven by sublimation of ices \citep{Snodgrass-MBCs-ISSI,Jewitt-MBCs-CometsIII}. 
Active asteroids include those whose tails are debris trails resulting from collisions, or other non-sublimation-driven mass loss (e.g., rotational disruption). The distinction between active asteroid and Main Belt comets is often difficult: the outgassing levels of these weakly active comets are so low that direct detection of the gas coma has only recently become possible using JWST \citep{Kelley-MBC-JWST}, and the categorisation generally depends on circumstantial evidence, such as repeated activity from orbit-to-orbit, or models suggesting long-duration activity rather than dust release in a single event.

Other observations that muddy the definition of comet vs asteroid in the Solar System include a number of apparently asteroidal bodies on comet-like orbits, as defined by \citet{Vaghi1973} and \citet{Levison1996}, which have been discovered in recent years \citep{Jewitt-MBCs-CometsIII}, and meteor streams with apparently asteroidal parent bodies \citep{Ye-CometsIII}. The limits of what can be described as cometary activity have been stretched by improved observation technology, revealing weak outgassing or dust release in objects previously thought to be inert. Examples of these include the asteroid Bennu, which was seen to be ejecting $\sim$cm-sized particles in {\it in situ} images obtained by the OSIRIS-Rex spacecraft, at an `activity' level that could never be detected remotely \citep{Lauretta-Bennu-activity}. 
Others, which have been called `dark comets', have shown no detectable activity by photometric or spectroscopic methods, but require non-gravitational acceleration that is best explained by the force of comet-like outgassing -- these include various near-Earth asteroids \citep{Seligman-dark-comets,Farnocchia2023}, and potentially the interstellar object 1I/`Oumuamua \citep{Micheli2018}.

Finally, while the word `comet' is generally used in reference to activity driven by sublimation of ices, it is important to remember that different chemical species sublimate at different temperatures: cometary activity near 1 au from the Sun is generally thought to be driven primarily by sublimation of water ice, but this is inefficient beyond 3 to 5~au heliocentric distance. Activity at larger distances is now regularly observed, with C/2014 UN271 and C/2017 K2 showing evidence for activity at 25--35~au \citep{Jewitt-K2,Bernardinelli-UN271}. In addition to these two well-known distant comets, C/2010 U3 and C/2019 E3 have also been observed showing activity beyond the orbit of Uranus \citep[i.e. beyond 19~au,][]{Hui2019, Hui2024}.
At these distances, activity is expected to be driven by more volatile ices, such as CO and/or CO$_2$ \citep{Meech-CometsII}. 

\begin{figure}
\centering
\includegraphics[width=1\textwidth]{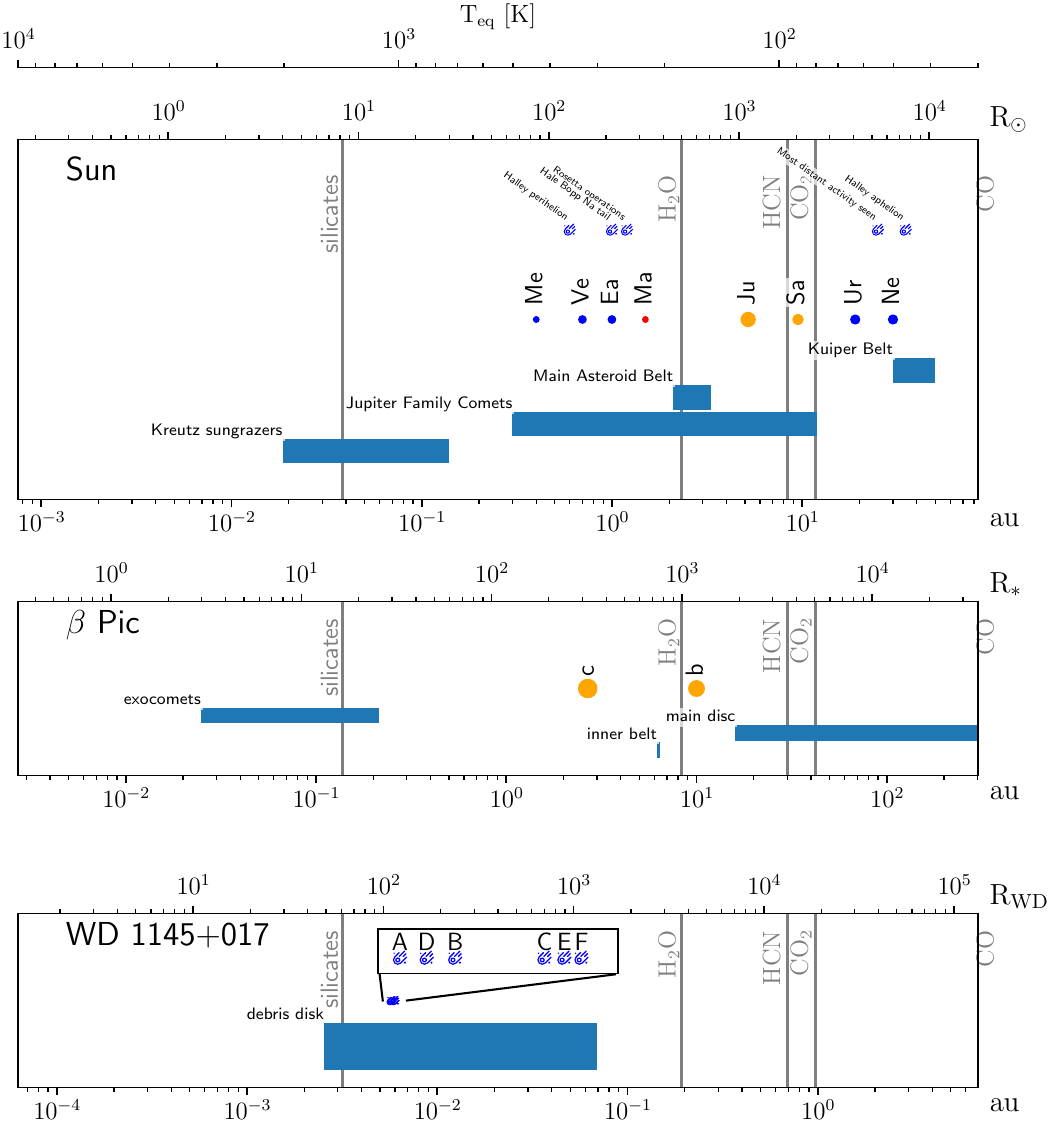}
\caption{The distribution of planets, main comet and asteroid belts around the Sun, and the equivalent analogues for Beta Pictoris and WD~1145+017.
All distances are logarithmic, scaled such that they all have the same equilibrium temperature for a blackbody, meaning that the various sublimation lines are at the same x-axis locations for comparison.
Distances from each star are labelled in stellar radii (upper scale) and astronomical units (lower scale).
For the Sun, Solar system comets are shown with notable associated distances, belts and planets.
For $\beta$~Pictoris, the planets and main belts are noted.
For the white dwarf WD~1145+017 the locations for the six transiting bodies are indicated, with a zoomed-in panel showing their labelled order from \citet{Vanderburg2015}.}\label{fig:SunBPic}
\end{figure}

At the other extreme, it is worth noting that any solid body will sublimate if heated strongly enough in a vacuum, and mass loss very close to the Sun does not require an icy composition. 
Asteroid (3200) Phaethon has been called a `rock comet' due to evidence for it losing material at its 0.14 au perihelion, which is not heated enough to sublimate most minerals, but will result in peak temperatures of $\sim 1000$~K that are sufficient to fracture rock and release dust in this way \citep{Jewitt-rock-comet,Ye-rock-comet}. 
It is likely that most of the Kreutz group of Sun-grazing comets, which are seen in coronographic observations,  fragment/disintegrate/sublimate at temperatures high enough to sublimate silicate rocks, at distances around 10 Solar radii \citep{Jones-near-Sun-ISSI}. 
Few near-Sun comets discovered by the Solar and Heliospheric Observatory (SOHO) actually survive perihelion; some of those that have been observed (when further from the Sun) have properties more typical of asteroids than `typical' icy comets \citep{Knight-322P}. 
In $\beta$\,Pic and WD\,1145+017, (systems observed to host exocomets) temperatures at the distances these comets are observed are comparable to the regime where Kreutz group comets are observed in our Solar System (see Fig.~\ref{fig:SunBPic}).
We must therefore be careful not to interpret the term `exocomet' as necessarily implying ice-driven sublimation or the presence of any icy component.

\subsection{Interstellar Objects}

An interstellar object is a macroscopic planetesimal from an extrasolar system observed passing through the Solar System on a hyperbolic trajectory. Three definitive interstellar objects have been detected to date: 1I/`Oumuamua in 2017 (originally 2017 U1 in \citealt{Williams17}), 2I/Borisov in 2019, and 3I/ATLAS in 2025. In this section, we provide a brief overview of these three objects and their divergent properties. For more detailed reviews see \citet{Fitzsimmons2023,Jewitt2023ARAA,MoroMartin2022,Seligman2023}. 

`Oumuamua and Borisov had distinctly contrasting properties. 1I/`Oumuamua displayed a myriad of bizarre properties mirroring both asteroids and comets, while 2I/Borisov displayed distinct cometary activity. 1I/`Oumuamua displayed no cometary tail in deep stacks of images \citep{Meech2017,Jewitt2017,Ye2017} or CO and CO$_2$ production \citep{Trilling2017}, but also a comet-like nongravitational acceleration \citep{Micheli2018}. It also displayed a reddened reflection spectrum \citep[i.e. it shows a positive slope toward the red part of the spectrum compared to the Sun's spectrum,][]{Bannister2017, Masiero2017, Fitzsimmons2017, Bolin2017}, a surprisingly low galactic velocity dispersion (inferred from the incoming trajectory kinematics) and inferred dynamical age \citep{Mamajek2017, Gaidos2017a,Hallatt2020,Hsieh2021}, and an elongated shape \citep{Meech2017,Bannister2017,Jewitt2017,Knight2017,Bolin2017,Drahus18,Fraser2017,Belton2018,Mashchenko2019}. There have been a wide range of hypotheses regarding the provenance of the object, including a fractal aggregate formed from diffusion-limited aggregation in the exterior of a protoplanetary disc, meaning it formed via similar processes observed on Earth that govern dust bunny and snowflake formation
  \citep{Moro-Martin2019}, to a comet-like object with low dust to gas ratio outgassing hypervolatiles \citep{fuglistaler2018,Sekanina2019,Seligman2020,Levine2021,Levine2021_h2,desch20211i,jackson20211i,Desch2022,Bergner2023}.

On the contrary, 2I/Borisov displayed definitive cometary activity associated with outgassing of volatiles \citep{Fitzsimmons2019,Jewitt2019b,deleon2019,Bolin2019,Guzik:2020,Hui2020,Mazzotta21}, more in line with the comets native to the Solar System. However, unlike typical Solar System comets 2I/Borisov was enriched in hypervolatiles, including a greater than unity ratio of CO to H$_2$O \citep{Cordiner2020,Bodewits2020}. It also exhibited a somewhat high degree of polarisation in dust in its outflow, implying that the object was more pristine and had not been significantly heated before \citep{Bagnulo2021,Halder2023}.

A third interstellar object was discovered on July 1 2025 \citep{Denneau2025} by the Asteroid Terrestrial-impact Last Alert System (ATLAS) survey \citep{Tonry2018a}, 3I/ATLAS. Preliminary observations revealed weak cometary activity and a reddened reflectance spectrum \citep{Opitom2025,Seligman2025,Alarcon2025,Jewitt2025,Kareta2025,Belyakov2025,Chandler2025,Marcos2025,Santana-Ros2025,Frincke2025,Jewitt2025b_HST,Puzia2025}. Moreover, pre-covery observations of 3I/ATLAS revealed that it was active in TESS and ZTF for $\sim2$ months prior to discovery, exterior to $\sim6$au heliocentric distances \citep{Feinstein2025,MartinezPalomera2025,Ye2025,Farnham_2025}. A variety of volatiles species have been detected actively being produced by the nucleus including CN \citep{SalazarManzano2025,Rahatgaonkar2025}, HCN \citep{Coulson2025}, CO$_2$ \citep{Cordiner2025,Lisse2025}, H$_2$O \citep{Xing2025}, and water ice \citep{Yang2025}.

\subsection{A definition of Exocomets}

The sporadic detection of transiting gas and dust clouds in the $\beta$ Pic (and other) system have been called `comet-like' since their discovery \citep{Ferlet_1987}. Following this earliest designation, the term `exocomet' to refer to these objects became common in the past decade. Another term found in the literature, \textit{falling evaporating bodies (FEBs)}, is now discouraged as it is misleading \citep{Strom2020} -- these objects are likely sublimating rather than evaporating, and are not necessarily in-falling. 

For the definition of the term ``exocomet", one needs to address the question of what can be designated by the term ``comet" in an extrasolar system: the requirement of orbiting a star other than the Sun is a direct consequence of the definition of an exoplanet \citep{Lecavelier_Lissauer_2022}.
Here, we propose to follow what is used in the Solar System since the earliest observations of comets in the sky: we propose restricting the use of the term `exocomet' to minor bodies that show signs of sublimation, along with their surrounding tails composed of dust and gas escaping the nucleus. The signs of a tail and/or a coma produced through sublimation processes can be detected either through their dust component via photometry or their gaseous component via spectroscopy.

Note that this definition is also in agreement with the etymology of the term ``comet", which comes from the Greek word `{\tt kom\^et\^es}', meaning `hairy star', which in turn comes from the word `{\tt kom\^e}' meaning `hair'.

This definition excludes planets that show a signature of atmospheric escape, such as HD209458\,b \citep{Vidal-Madjar_2003}, as they are not minor bodies. In the same spirit, debris discs can be supplied by exocomets and be mostly composed of remnants of exocometary material, and are therefore of prime interest in the exocomets studies; however, they are not considered as ``exocomets" per se as they cannot be directly linked to individual bodies. On the other hand, sublimating bodies observed in the light-curves of polluted white-dwarves, which generally show distinct asymmetry by the transiting cometary tail (e.g. \citealt{Vanderburg2015, Vanderbosch2021}) do fall within this definition of an exocomet.

Similarly to comets in the Solar System, the term ``exocomet" is thus used in reference to activity driven by sublimation of ices or rock at short distance to the parent star (Sect.~\ref{Solar System comets}). Fortunately, in extrasolar systems, we do not face the same difficulties as in the Solar System where the question arises of what can be described as cometary activity in the context of the detection of very weak outgassing or dust release from objects previously classified as asteroids. In fact, in extrasolar systems we do not detect the nuclei but the result of the sublimation process by observing the comae and tails, making the detection and the classification a single joint problem.   

In the context of white dwarfs, we use the term exocomets to describe any minor bodies with a tail composed of dust and/or gas, regardless of its period and eccentricity. In those systems, the outgassing mechanism is less clear, and it could be driven by sublimation, collision, tidal disruption, or a combination of several mechanisms.

The eccentricity of the orbit in the definition of the term ``exocomet" is irrelevant, as in many cases we are not able to determine their eccentricity and they will be considered exocomets regardless of it.
Most of those observed to date have close periastron passages, resulting in temperatures that lead to sublimation of rocky material. 
Those observed in spectroscopy give access to their radial velocity at the time of the transit. The measured velocities can range from zero up to hundreds of kilometres per second, both blue-shifted and redshifted, indicating motion on a variety of orbits and placing them typically at very short distances to the parent stars. Note here that the comet nucleus does not need to be in-falling (i.e., grazing or hitting the star) to be sublimating. Indeed, we have thousands of observations of Kreutz sungrazers in the Solar System, but have yet to convincingly detect a `Sun Diver' \citep{Jones-near-Sun-ISSI}. 

Finally, it is worth mentioning that \cite{Strom2020} proposed a definition of the term exocomet as: \textit{``In this paper, we use the word `exocomet' to describe comets which orbit other stars than the Sun and which exhibit some form of observable activity such as the release of gas or dust, e.g. through a coma or tails of ions or dust.''} The definition proposed here is in overall agreement with that of \cite{Strom2020}. However, interstellar objects are called `exocomets' in \cite{Strom2020}. This is not consistent with the condition of orbiting stars other than the Sun, and therefore, in this work we do not consider interstellar objects as exocomets. We have expanded on the existing definition by addressing potentially debatable cases such as sublimating exoplanets displaying an asymmetric exocomet-like transit, for instance. This work aims to draw further constraints on what is and what is not considered an exocomet throughout this series of manuscripts on exocomets.

\section{Exocomet Observations}\label{sec:obs}

In this section, we will first talk about the detection of individual exocometary `bodies', which is what we previously defined as `exocomets'. Next, we provide a brief overview of the detection of exocometary `material' attributed to the remnants of exocomet disintegration or sublimation in debris discs and around white dwarfs.

\subsection{Detection of exocometary `bodies'}\label{sec:bodies} 

\subsubsection{Spectroscopy}\label{sec:spec} 

The first observations of exocomets were made in the 1980's through spectroscopy. Their violent sublimation was observed for the first time close to the exoplanet host star $\beta$~Pictoris as variations in the Ca\,{\sc ii} K absorption line \citep{Ferlet_1987}. The stellar nature of the variations was ruled out as the variable features appeared superimposed on photospheric lines. Incidentally, $\beta$~Pictoris is a rapidly rotating star, with very broad and deep photospheric lines. The features produced by exocomets are instead narrow and often shallow, with small equivalent widths (EW), indicative of the small amount of gas included in the tail and/or coma of the sublimating body. Within estimated distances of $\lesssim$0.1\,au \citep{Beust1989, Beust1990}, the rocky material sublimates, creating a cloud of atoms that rapidly photo-ionises, including strongly absorbing calcium ions. Strong, stochastic absorption by ionised calcium has since become a land-mark property of exocomets observed through transit spectroscopy in other systems. In $\beta$\,Pictoris, exocomet events are very frequent, and many thousands of events have been observed spectroscopically \citep[e.g.][]{Kiefer2014, Tobin2019}. Exocomet transits usually last in the order of hours or days (\citealt{Kiefer2014}, \citealt{Lecavelier22}) and vary over time. However, these events are inherently stochastic. They are not predictable, it is virtually impossible to determine whether any of these exocomets have transited more than once, and their orbits can generally be constrained only poorly. In some cases, gravitational acceleration can be observed for comets during very close periastron passages observed with high-resolution time-series spectra, giving some constraint on their orbital characteristics \citep{Kennedy_2019}. Generally, their orbits are assumed to be nearly parabolic, allowing the periastron distance to be constrained together with its orientation (argument of periastron). In $\beta$\,Pictoris, this has pointed to the existence of dynamical families of bodies \citep{Kiefer2014}, and such exocomet censuses can thus be used to infer properties of the dynamical evolution of exocomet-hosting systems \citep{Beust2024}.

The mechanisms of sublimation of rocky material into atoms and subsequent interactions with the strong radiation field (and/or wind) of the star are complicated physical processes that require both detailed observations and theoretical modelling to allow them to be accurately described (e.g. \citealt{Karmann2003}; \citealt{Vrignaud2025}). Although the observed radial velocity of the ionized calcium clouds typically vary between blue and redshifts of several tens of km/s, theory predicts that the lines may be expected to be blueshifted by hundreds of km/s, because calcium ions get efficiently accelerated due to radiation pressure \citep{Beust1989}. The relatively modest observed radial velocities can be reproduced by tracking the dynamics of dust particles sublimating from the nucleus and releasing ions upon sublimation \citep{Beust1990}. The density of ions is greatest at a shock-front that forms between the nucleus and the star. In the shock, the opacity is high enough for self-shielding to occur, tempering the accelerating effect of radiation pressure; and causing the majority of the observed absorption. Collisions with unseen volatiles (that do not feel strong radiation pressure) may also play a role in slowing down the ions \citep{Beust1990}. As the shock is co-moving with the nucleus, the observed Ca {\sc ii} absorption lines effectively trace the orbital velocity of the nucleus.  

A crucial property of transiting exocomets is that their clouds are spatially confined close to the sublimating nucleus, which can be observed thanks to the Ca {\sc ii} H \& K doublet. In the case of an optically thin cloud ($\tau \lesssim 1$), both lines are unsaturated and the ratio between the line depths tends to the ratio of the oscillator strengths, equal to two. In the case of a moderately optically thick cloud ($\tau \gtrsim 1$), the Ca {\sc ii} doublet lines are partially saturated, making the line ratio approach unity. If the cloud is confined, it does not fill the entire stellar disc, and so the saturated absorption lines will not entirely attenuate the star. This means that the fill-fraction of the transiting cloud can be constrained, providing direct observational evidence of its spatial confinement \citep{Lagrange1989,Vidal-Madjar_1994,Kiefer2014,Vrignaud2024a}. This has been driving evidence for the interpretation that these lines are caused by exocomets in $\beta$ Pic as well as in other systems \citep[e.g. HD\,172555,][]{Grady2018}.

Spectroscopic observations probe the gaseous component of exocomet comae and tails and can provide a unique insight into the composition of exocomets as absorption lines detected during transit can reveal the different species (typically ionized) present in this gas. Although it is not possible to measure the total amount of sublimated refractory material of an exocomet given that we are only able to observe the portion of gas crossing the line of sight of the star, it is possible to obtain the column densities of different species and estimate relative abundances. New methodologies such as a curve of growth (used in interstellar medium studies) adapted to transiting exocomets can also reveal the excitation temperature of the gas tail along with column densities and stellar fraction covered by the transit \citep{Vrignaud2024a, Vrignaud2025}.


Such transient and variable features attributed to exocomets have not yet been detected around white dwarfs through spectroscopy. However, spectroscopic studies have uncovered circumstellar gas in orbit around white dwarfs and evidence of exocometary material accreting onto their atmospheres (see Sec.~\ref{sec:material} for more details). 

\subsubsection{Photometry}\label{sec:photo}

Exocomets also release dust that is lost with the sublimation of volatiles, and this dust can transit the star in the same way as the gas. Because of this, precise long-term photometric monitoring has allowed the identification of broad-band exocomet transits as well \citep{Zieba2019}. Photometric monitoring of transits allows direct detection of the geometrical extent and the optical thickness of the passing dust cloud; therefore, coupled with models of dust production, this allows an estimate of the dust production rate and the size of the comet's nucleus at the origin of the transiting dust tail. In the case of $\beta$~Pictoris, statistics of photometric exocomet transits point to a nucleus size distribution that is similar to that of small bodies in the Solar system, set by collisional equilibrium \citep{Lecavelier22}.

Detecting exocomets in photometry focuses on the transit shape of the observations. An important characteristic in most of the light curves of an exocomet transit is the asymmetry caused by the cometary tail passing the line of sight after the nucleus. As an exocomet passes the star, the stellar flux will decrease steeply, characteristic of the coma, followed by a slow increase back to the full stellar flux levels due to the optically thinner tail occulting the star \citep{lecavelier1999,lecavelier1999a}. Symmetric exocomet transit light curves are possible (i.e.: when the cometary tail is along the line of sight), but in this case it is almost impossible to distinguish them from other symmetric transits such as binaries and exoplanets \citep[][]{lecavelier1999}. The exact shape of the exocomet transit depends on the properties of the host star, the effects of radiation pressure and stellar winds on the sublimation rate of the exocometary body, and the exocomet body itself i.e: its orbital geometry, size, etc. \citep{Strom2024}. Variable and asymmetric transits are also the main signature indicating the presence of exocomets around white dwarfs \citep{Vanderburg2015}. In addition, while the periodicity has been determined for some systems, it appears that many of the systems have no periodicity at all \citep{Bhattacharjee2025}. Around main-sequence stars, there are two probable cases of periodic exocomet events in photometry. One is the transit of a string of exocomets in front of KIC\,8462852 with potential periods of $\sim$2~years or 928~days \citep[respectively]{boyajian16,Kiefer_2017}. The other one is a possible explanation for the three transits seen in KIC\,3542116 that had a periodicity of $\sim$ 92 days, although several expected transits for this system were not observed likely due to differences in transit depth \citep{Rappaport2018}.

Estimating the physical parameters of exocomet bodies just from the transit lightcurve is challenging, as these events are rare, sporadic in nature, and so far observed from space-based instruments (and therefore only observed in a single photometric band), limiting the ability to constrain properties such as dust reddening. However, various modelling approaches have been developed to characterise some physical parameters that can be derived. \cite{lecavelier1999} first developed numerical exocomet models to infer parameters such as the dust distribution and production rates of cometary nuclei. Further numerical approaches have been developed since, introducing multiple wavelengths \citep{Kalman2024}, and using Monte Carlo methods \citep{Lukyanyk2024} to refine their models. However, numerical models also require detailed knowledge of the host system and rely on assumptions between model parameters that remain uncertain, such as the relationship between dust production and the distance to the star.

A simpler alternative focuses on the key characteristic of exocomet transits: their asymmetric profile. This empirical approach to fitting transit lightcurves is also beneficial for initial detections before attempting detailed physical parameter estimations. Several empirical models have been developed, particularly for recent photometric surveys, and include modified planet transit fits \citep{Zieba2019}, and modified exponential fits to the data to characterise asymmetry \citep{Rappaport2016, Kennedy_2019, Lecavelier22,Norazman2025}. While empirical fits do not provide much physical information about the cometary bodies, these are simpler and are valuable for identifying new candidates, which can then be characterised in detail with the numerical models.

\subsection{Detection of exocometary material} \label{sec:material}

This section refers to the detection of material (gas and dust) attributed to be remnants of exocomets as detected in circumstellar discs, rather than the direct detection of exocomets as previously defined.
The presence of dust and gas in mature planetary systems, where the star has reached the main sequence, can in some cases (such as that of $\beta$~Pictoris; \citealt{Lecavelier_1996}) be attributed to a secondary origin. This material is often referred to as ``exocometary", as one of the hypothesis is that it was generated through sublimation of comets as they approach the central star, although different mechanisms such as collisional cascades \citep{Artymowicz1997,Wyatt2008} have been raised, especially to explain the material at longer distances from the star, where sublimation might not be as effective. 
But the presence of gas even at long distances from the star is hard to explain through collisions, and it requires bodies with a high level of ices. Therefore, outgassing seemed a more likely explanation \citep{Lecavelier_1996,Matra2017b,Kral2019}, where cometary-like events are thought to be a major contributor. 

Over the years, detections of stable gas components across different wavelengths have been made for various chemical species. Some of the first detections were obtained in Ca{\sc ii} and Na{\sc i} \citep{Hobbs1985,Vidal-Madjar_1986}, and in the Fe{\sc ii} lines in UV \citep{Kondo1985}. 
In addition to the detection of a large number of refractory elements around main-sequence stars (e.g., Mg{\sc ii}, Al{\sc ii}, Si{\sc ii}, Cr{\sc ii}, Zn{\sc ii}), the UV spectroscopy allowed also the discovery of molecular CO gas around $\beta$\,Pictoris \citep{Vidal-Madjar_1994,Jolly1998, Roberge2000}. A detection of H$_2$ was absent in FUSE far-UV spectroscopic observations \citep{Lecavelier_2001}, with a CO/H$_2$ ratio several orders of magnitude larger than that observed in the interstellar medium. In these conditions, the CO is not protected from the destructive interstellar UV radiation field and has a lifetime of about 120~years \citep{Dent_2014}, demonstrating its origin from the sublimation of icy bodies. This scenario is strengthened by the measured temperature of CO at $\sim$30\,K and the high $^{13}$CO/$^{12}$CO isotope ratio revealed by UV spectroscopy and likely due to the fractionation during the sublimation process \citep{Jolly1998}. The advent of ALMA\footnote{Atacama Large Millimeter/submillimeter Array \url{www.almaobservatory.org/}} confirmed the presence of CO around $\beta$\,Pictoris \citep{Dent_2014}. To date, we know between 20 and 30 main-sequence stars that show evidence of volatile gas, which is attributed to a secondary origin through outgassing of comets \citep[e.g.][and references therein]{Rebollido2022, Moor2019}.
The expected evolution of stars and their planetary systems, according to current planetary formation theories, implies that as the system settles, the activity of outgassing and colliding bodies decays with time, until the star leaves the main sequence and a second-generation debris disc (i.e. a compact disc assumed to come from collisions or disruption of small bodies) remains \citep{wyatt2015}. 

Exocometary material has been detected around white dwarfs too, via atmospheric pollution and debris disc. \citet{Veras2024} provides a recent review on different processes that bring extrasolar minor bodies close to the tidal radius of the white dwarf. Polluted white dwarfs, which exhibit elements heavier than hydrogen in the atmosphere, are a common occurrence, and they are a result of accretion of exoasteroids and exocomets. 
Sometimes, it is possible to derive the chemical compositions of the exo-minor bodies because multiple species (both refractory and volatile elements) are detected in the atmosphere of the white dwarf \citep[e.g.][]{Rogers2024a}.  Spectroscopic observations of polluted white dwarfs uniquely return the chemical composition of the exoplanetary material, the diversity of which is discussed in \citet{Xu2024b}. The most heavily polluted white dwarfs often display an infrared excess from a debris disc (see \citealt{Malamud2024} for a recent review on white dwarf debris discs). Circumstellar absorption and emission lines have been detected around two dozen white dwarfs, and they are typically attributed to ongoing tidal disruption events \citep[e.g.][]{Gaensicke2006, Debes2012b}. These circumstellar lines were first detected in the Sloan Digital Sky Survey (SDSS, \citealt{Gaensicke2006}). They can be variable on the hourly, weekly, and monthly timescales; some display periodic variations, while others appear to be more stochastic \citep[e.g.][]{Manser2016, Xu2019a}. 
 The lines are double-peaked, consistent with a rotating gas disc. The strongest lines tend to be the calcium infrared triplet around 8500~{\AA} and the width can be up to 1000~km/s, making them easily detectable with low-resolution spectroscopy. Most of the emission line systems are detected in SDSS and spectroscopic observations of polluted white dwarfs \citep{Dennihy2020b, Gentile-Fusillo2021a}. On the other hand, circumstellar absorption lines can be much narrower and therefore hard to detect. High-resolution spectroscopy (resolving power $>$ 10,000) is often needed to detect those systems. 

\section{Census of Exocometary Systems}\label{sec:census}

In this section, we provide an inventory of all the reported detections to date of exocometary system candidates. We present all exocomet detections via spectroscopy and photometry, including photometric detections around white dwarfs. For completeness, although they are not technically considered exocomets, we also give an overview of detections of interstellar objects.

\begin{figure}
\centering
\includegraphics[width=1\textwidth]{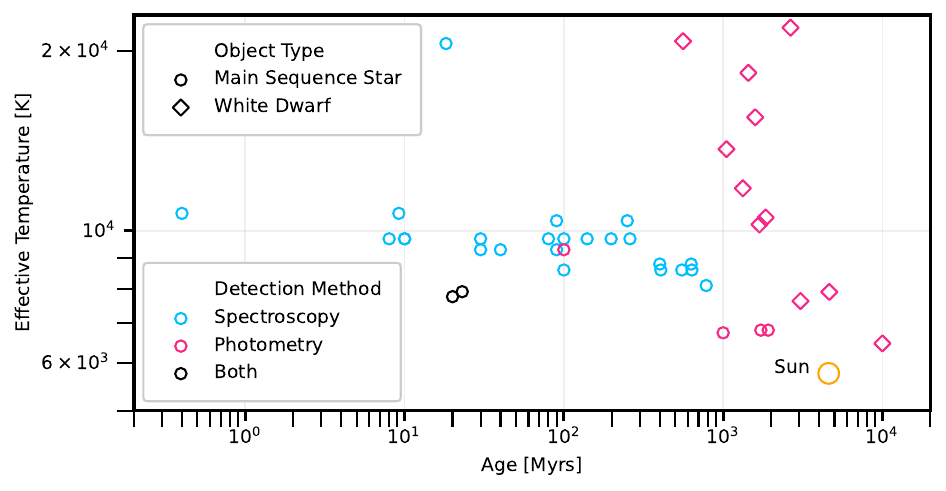}
\caption{Overview of ages and T$_{\rm eff}$ for stars with reported exocomet detections for main sequence and white dwarf systems. Stellar T$_{\rm eff}$ have been adopted from the spectral types in Tables \ref{tab:exocomet_candidatesSP} and \ref{tab:exocomet_candidatesPH} using the conversion from \cite{Pecaut_Mamajek2013}. Unclear and false positive detections are excluded from the figure. White dwarf objects are also listed in Table~\ref{tab:exocomet_candidatesPHWD}.}\label{fig:detections}
\end{figure}

\subsection{Spectroscopic detections}

Most spectroscopic detections come from individual campaigns targeting stars with evidence of circumstellar material, such as debris discs. In particular, $\beta$~Pictoris has been intensively studied since the first detection of exocomets around the star. 
Most of the other discoveries were made in small searches with few targets \citep[e.g.][]{Montgomery2012,Montgomery2017}, with the exception of three large surveys: (1) \cite{Iglesias18}, which was focused on the gaseous component of debris discs; (2) \cite{Rebollido2020} which included a sample of stars with evidence of previous circumstellar material, and (3) \cite{BendahanWest2025} which represents so far the only unbiased search for exocomets, with a partially automated analysis of the full HARPS-S archive. To date, variable absorption features attributed to exocomet transits have been reported for 30 systems. Most of these spectroscopic detections have been around A-type stars and a few late B-type stars. This is likely due to their rotationally broadened spectral lines which facilitates the detection of the superimposed narrow absorption features produced by gas transiting the line of sight of the star. However, this observational bias does not imply the lack of exocometary activity in other types of stars as, for instance, we know of the existence of comets around the Sun, a G-type star. Another bias seem to be the systems being observed close to edge-on, as shown by the results of \cite{Iglesias18} and \cite{Rebollido2020}, and the canonical exocomet host $\beta$~Pictoris.

It is worth noting that we cannot be sure that all detections possibly attributed to exocomet transits are in fact due to exocomets. Cases such as that of $\beta$~Pictoris and HD\,172555 are considered as ``confirmed'' exocomets as they have been detected in both photometry and spectroscopy and through several different chemical species. On the other hand, there are many cases with a single or few detections which remain unconfirmed so far. Therefore, we refer to them as ``candidates'' until further confirmation. 
All candidate and confirmed exocomet-host systems are presented in Table~\ref{tab:exocomet_candidatesSP} and their ages and T$_{\rm eff}$ distribution are shown in Fig. \ref{fig:detections}. However, for two of these objects; $\alpha$ Pictoris and HD\,138629, the detections were reported as inconclusive and could be either attributable to exocomets or interstellar gas \citep{Hempel2003, Lagrange-Henri1990}. Two other cases, that of HD\,256 (HR\,10) and $\phi$ Leo, were initially considered as exocometary systems, until further investigation revealed a different scenario. In the case of HR\,10, the variations were due to the system being a binary and, in the case of $\phi$ Leo, the variability was the result of $\delta$~Scuti pulsations \citep{Montesinos2019, Eiroa2021}. Nevertheless, all systems are listed in Table~\ref{tab:exocomet_candidatesSP} for completeness.\\

\input{TableSpec}

\subsection{Photometric detections}

The detection of exocomets with photometry became possible with the \textit{Kepler} space telescope. The first interesting photometric event was pointed out by \cite{boyajian16} in the \textit{Kepler} light curve of KIC\,8462852. The system exhibits irregular asymmetric transits, where one of the more attractive scenarios may be explained as the sublimation of a string of exocomets with a dust production rate similar to Hale-Bopp \citep{Schleicher1997, Jewitt_Matthews1999}. 
A more complete inspection of the \textit{Kepler} data via a visual survey allowed the finding of asymmetric transits around two systems: KIC\,3542116 and KIC\,11084727 \citep{Rappaport2018}. The light curves matched the predictions of \cite{lecavelier1999}. In addition, an automated search of the same dataset was conducted in \cite{Kennedy_2019} to follow up on the two discoveries. They recovered the transits with their pipeline and also identified a potential third system to host exocometary activity; KIC\,8027456. They obtained a detection frequency of $3\times10^{-5}$ for transits deeper than 0.1\% for the survey consisting of $\sim$ 150,000 stars. Furthermore, \cite{Kennedy_2019} concluded that, despite an unbiased \textit{Kepler} dataset (compared to the biases faced for spectroscopic surveys), circumstantial evidence suggests that there may be a real tendency of exocomet detections around young, early-type stars.  

The natural extension of the \textit{Kepler} mission was the Transiting Exoplanet Survey Satellite (TESS), where exocomets have been detected around $\beta$ Pic \citep{Zieba2019}, making it the first star to have exocomet detections in both photometry and spectroscopy. Further analyses of the available TESS data around $\beta$ Pic have provided multiple more exocomet detections \citep{Pavlenko2022,Lecavelier22}. With up to 30 exocomet transits identified, a statistical analysis allowed for a size distribution of the cometary nucleus to be estimated \citep{Lecavelier22}. More recently, observations around the extreme debris disc system RZ Psc with TESS also identified 24 exocomet transits, allowing for a second system with sufficient detections to infer their size distribution \citep{Gibson2025}. On a larger scale, statistical searches for exocomets have been conducted to determine their frequency with relation to their host star properties in both \textit{Kepler} and TESS (\citealt{Kennedy_2019, Norazman2025}), yielding candidates seen in Table~\ref{tab:exocomet_candidatesPH}. While \cite{Norazman2025} recovered the $\beta$ Pic detection and identified a new F-type main-sequence candidate, they also detected signatures around G-type main-sequence stars and two detections around giant stars. The two detections around G-type main-sequence stars suggest that we could also detect exocomets around later-type stars, and it may be that the young, early-type detections were observational effects. However, the number of overall exocomet detections in photometry is still too low to obtain robust statistics. Other instruments have also been used to search for exocomets, such as CHEOPS \citep[ESA's CHaracterising ExOPlanet Satellite;][]{Benz2021}, which potentially detected a transit around HD\,172555 \citep{Kiefer2023}. The key outcome from searches in photometry is that exocomets are rare, and the rarity of these events makes it difficult to characterise them and form strong statistical conclusions. Therefore, more detections of exocomets with high-precision photometry, such as the upcoming PLATO mission, could help address some of these issues, but validating single transits remains a challenge. The exocomet candidates in photometry are shown in Table~\ref{tab:exocomet_candidatesPH}.

The first detection of a transiting exocomet around the white dwarf WD~1145+017 was found using the extended \textit{Kepler} mission (K2, \citealt{Vanderburg2015}). The K2 data show six stable periods between 4.5--4.9~hr, which is within the tidal radius of the white dwarf. Follow-up studies show that the transits are asymmetric, deep, and variable \citep{Rappaport2016}. However, the dust production mechanism is unclear -- it could be sublimation, tidal disruption, collision, or a combination of several scenarios. In addition, WD~1145+017 also has a heavily polluted atmosphere, displays variable circumstellar absorption lines, and has an infrared excess from a dust disc. The emerging picture is that WD~1145+017 might represent the early stage of a tidal disruption event, with the exocomets still in the process of actively disintegrating. The WD~1145+017 is shown in Fig. \ref{fig:SunBPic} for comparison with the Solar System and $\beta$~Pictoris. Follow-up analysis considers a sample of 1148 white dwarfs observed by K2 and did not identify any new transiting objects, placing an upper limit of 12\% on the presence of disintegrating bodies \citep{vanSluijsVanEylen2018}.

The Zwicky Transient Factory (ZTF) has been very successful in finding more transit bodies around white dwarfs due to its long observing baseline and faint magnitude limit. So far, ZTF has identified nine new systems, each displaying unique properties, as summarized in Table~\ref{tab:exocomet_candidatesPHWD}. These systems are also shown in Fig. \ref{fig:detections} along with the main sequence stars candidates. TESS has found quasi-continuous transits around WD~1054-226 with a predominant period of 25.02~hr \citep{Farihi2022}. Otherwise, TESS found no new transiting systems within a sample of 313 polluted white dwarfs \citep{Robert2024}.

\input{TablePhoto}

\input{TablePhotoWD}

\subsection{Census of Interstellar Objects}

The physical properties of the first two detected interstellar objects, 1I/`Oumuamua and 2I/Borisov, are summarized in Table 2 of \cite{Jewitt2023ARAA}. 
The radii of 1I/`Oumuamua and 2I/Borisov are of order $\sim80$ m \citep{Jewitt2017,Meech2017,Drahus18,Knight2017} and $\sim400$ m \citep{Jewitt2020}. The diameter of the nucleus of 3I/ATLAS is uncertain given the faint cometary activity, with an upper limit of $\sim20$km \citep{Seligman2025,Loeb2025}.
Lightcurve inversion suggests 1I/`Oumuamua had a 6:6:1 oblate shape \citep{Mashchenko2019}, while the shapes of 2I/Borisov and 3I/ATLAS were unconstrained due to their comae. 
1I/`Oumuamua and 2I/Borisov both  exhibited comet-like nongravitational acceleration \citep{Micheli2018}. The non-gravitational acceleration of 3I/ATLAS are currently unconstrained at the time of the writing of this manuscript.

It is challenging to directly compare the physical properties of interstellar objects with those of exocomets, even though their nominal size distributions are consistent with those of exocomets around $\beta$ Pic \citep{Lecavelier22}. This preliminary similarity should be interpreted with caution, as `Oumuamua did not display cometary activity, whereas Borisov and ATLAS did.
It is impossible to trace an interstellar object to a progenitor system due to the chaotic nature of stellar orbits in the galaxy and the uncertainties in phase-space measurements \citep{Zhang18}. 
However, the incoming kinematics of an interstellar object trajectory can be used to give a rough dynamical age.  1I/`Oumuamua had a surprisingly low  velocity (26 km s$^{-1}$) compared to the local standard of rest which implies  a dynamical age of $\tau \sim$ 100 Myr which was pointed out  by \cite{Mamajek2017} and \cite{Gaidos2017}.  Numerical integrations of many realizations of  galactic trajectories demonstrated that the object was  most likely associated with the local Orion Arm and co-moving with Carina or Columba stellar associations \citep{Hallatt2020}. The same study also pointed out that the larger velocity of 2I/Borisov (32 km s$^{-1}$) is consistent with an age of  $\tau \sim10^9$ yr post ejection \citep{Hallatt2020}. \citet{Hsieh2021} argued that the lower dispersion of 1I/`Oumuamua was evidence for formation in the core of a giant molecular cloud, because of their low velocity dispersion. 
This would be consistent with the hypothesis that `Oumuamua's bizzarre properties could be attributed to it being composed of primarily H$_2$ ice \citep{Seligman2020}. However, there are two caveats to the assumption that interstellar objects trace the stellar kinematics that are worth noting. First, \citet{Forbes2024} demonstrated that the galactic background of interstellar objects bears more similarity to stellar streams in nature, given that a single star can eject a large number of planetesimals.  Secondly, there should be a contribution to the galactic population from stars that have already died \citep{Hopkins2023,Hopkins2024}. Despite these caveats, it appears that the only conclusion regarding the progenitor systems of 1I/`Oumuamua and 2I/Borisov is that 2I/Borisov likely originated from a much older system than 1I/`Oumuamua. 3I/ATLAS had significantly larger excess velocity than 1I/`Oumuamua and 2I/Borisov (58 km s$^{-1}$), consistent with a $\sim3-11$ Gyr age \citep{Taylor2025} and possibly a  thick disk origin \citep{Hopkins2025b}.

The composition of 1I/`Oumuamua is somewhat unconstrained because limited observations of the object did not recover any gas production. For complete tables describing the entirety of the compositional measurements, production rates and upper limits of various volatiles species of 1I/`Oumuamua and 2I/Borisov during their apparitions, see Tables 3 - 5 in \cite{Jewitt2023ARAA}. For 1I/`Oumuamua upper limits on the production of CN, C$_2$, C$_3$ \citep{Ye2017}, OH \citep{Park2018}, CO$_2$ and CO \cite{Trilling2018} are given. 

2I/Borisov had an apparition that was observable for several months. Detailed spectroscopic measurements were made characterizing the production rates of multiple species pre- and post-perihelion. Measurements of CO post-perihelia \citep{Bodewits2020,Cordiner2020} indicated that the object was enriched in CO with respect to H$_2$O \citep{Xing2020} which is atypical of comets in the Solar System \citep{Bockelee17, McKay2019,Seligman2022PSJ}. The composition of 3I/ATLAS is unconstrained at the time of writing this manuscript.

\section{Summary}\label{sec:summary}

This manuscript provides a general overview of exocomets, addresses the definition of what we consider as ``exocomets" throughout this collection of works on exocomets, and gives a summary of all the reported detections to date. Below, we summarize some of the key points covered in this overview:
\begin{itemize}
    \item The definition of ``exocomet" refers to minor bodies in extrasolar systems that have a coma and/or tail composed of dust and gas, a product of sublimation close to the star. This definition does not include gas or dust found in continuous distributions orbiting around stars or white dwarfs, or material produced in collisions in debris discs, or pollution in white dwarf's atmospheres.

    \item To date, the system $\beta$~Pictoris remains as the premier exocomet host with thousands of detections of exocometary transits. Thanks to its proximity to Earth and its very high exocometary activity it has allowed us to study the dynamics and composition of exocomets better than any other system. The best white dwarf analogue is WD~1145+017, which hosts numerous transits and a debris disk.

    \item Spectroscopic and photometric detections of exocomets allow us to estimate different physical parameters of exocomets. While spectroscopic observations give us information regarding the composition and dynamics of exocomets, photometric observations allow us to constrain parameters such as the size of exocomets nuclei and dust production rates.

    \item Exocomet candidates have been reported in about 40 stars, $\sim$30 of them through spectroscopy and 11 via photometry. Only two stars, $\beta$~Pictoris and HD\,172555 have detections both in photometry and spectroscopy. In addition, signs of exocomet transits have been identified in 11 white dwarfs via photometry. Most of the candidates so far are around stars and white dwarfs younger and hotter than the Sun. However, it is worth noting that these statistics might be biased by detection constraints, thus the true distribution of exocomets hosts remains unknown. 

\end{itemize}

This overview of exocomets is part of a collection of manuscripts studying different aspects of exocometary science. Korth et al. (submitted) describes observations of exocomets in more detail, Vrignaud et al. (submitted) studies the physical processes of exocomets, Lu et al. (submitted) is devoted specifically to the $\beta$~Pictoris system, \cite{Bannister2025} discusses the origins and initial reservoirs of exocomets, Mustill et al. (submitted) addresses the evolution and end states of exocomets, and Lecavelier \& Str{\o}m (submitted) proposes a nomenclature for naming individual exocomet detections.







\backmatter



\bmhead{Acknowledgements}

We gratefully acknowledge support by the International Space Science Insitute, ISSI, Bern, for supporting  and hosting the workshop on ``Exocomets: Bridging our Understanding of Minor Bodies in Solar and Exoplanetary Systems'', during which this work was initiated in July 2024.

I.R. acknowledges support from the Research Fellowship Program of the European Space Agency (ESA).

D.Z.S. is supported by an NSF Astronomy and Astrophysics Postdoctoral Fellowship under award AST-2303553. This research award is partially funded by a generous gift of Charles Simonyi to the NSF Division of Astronomical Sciences.  The award is made in recognition of significant contributions to Rubin Observatory's Legacy Survey of Space and Time. 

S.X. is supported by NOIRLab, which is managed by the Association of Universities for Research in Astronomy (AURA) under a cooperative agreement with the National Science Foundation.

A.N. is supported by the University of Warwick and the Royal Society.

D.I. acknowledges support from the Science and Technology Facilities Council via grant number ST/X001016/1.

A.L. acknowledges support from the Centre National des Etudes Spatiales (CNES). 

M.T.B. appreciates support by the Rutherford Discovery Fellowships from New Zealand Government funding, administered by the Royal Society Te Ap\={a}rangi.

H.J.H is supported by funding from eSSENCE (grant number eSSENCE@LU 9:3), the Swedish National Research Council (project number 2023-05307), The Crafoord foundation and the Royal Physiographic Society of Lund, through The Fund of the Walter Gyllenberg Foundation.

C.S. acknowledges support from the UK Space Agency and Science and Technology Facilities Council.


\section*{Declarations}

\bmhead{Conflict of interest/Competing interests} Not applicable.



\bigskip





\begin{appendices}






\end{appendices}


\bibliography{sn-bibliography}

\end{document}

%% file: TableSpec.tex
\begin{table*}
\begin{center}
\caption{Stars with reports of variable absorption features, most typically in the Ca\,{\sc ii} K line (among others), attributed to exocometary activity. This Table was built upon Tables 1 and 2 in \cite{Strom2020}.} 
\label{tab:exocomet_candidatesSP}
\begin{tabular*}{1.0\textwidth}{@{\extracolsep\fill}llcccr} %
Name & Other ID & Sp. Type & Age [Myrs] & Species & Reference      \\
\hline
\multirow{ 3}{*}{$\beta$~Pic*} & \multirow{ 3}{*}{HD\,39060}  & \multirow{ 3}{*}{A6V} & \multirow{ 3}{*}{23$\pm$3 [1]} & Ca, Na, Fe, Mg, S  & [2, 3, 4,   \\
 & & & & Al, C, Co, Zn & 46, 47, 51 \\
  & & & & Ni, Cr, Mn, Si & 48, 49, 50] \\
HD\,172555*    &    	HR\,7012   & A7V & 20$\pm$5 [5] & Ca, Si, C, O  & [6, 7, 46]       \\
49\,Cet & HD\,9672           & A1V & $\sim$40 [10] & Ca, C & [8, 9, 10]       \\
HD\,21620   &   HR\,1056    & A0V & $\sim$80 [11] & Ca   &  [12]       \\
HD\,32297   & HIP\,23451  & A0V & $\sim$30 [13] & Na  &  [14]       \\
HD\,37306 & HR\,1919         & A1V  & $\sim$30 [15]  &  Ca, Fe, Ti   &  [16]      \\
HD\,42111 & HR\,2174        & A3V & 406$\pm^{141}_{177}$ [17] & Ca, Fe & [12, 18]   \\
$\lambda$~Gem & HD\,56537   & A3V & $\sim$550 [19] & Ca & [20]       \\
HD\,58647      & HIP\,36068  & B9IV   & 0.4$\pm$0.1 [21] &  Ca &  [20]       \\
$\phi$~Gem & HD\,64145      & A3V & 637$\pm^{111}_{199}$ [17] & Ca & [20]       \\
$\beta$~Car & HR\,3685      & A2IV & $\sim$400 [23]  & Ca, Na & [22, 23]   \\
HD\,85905        &  HR\,3921  & A2V & $\sim$630 [23] & Ca &  [24, 23]    \\
$\delta$~Crv & HD\,108767   & A0IV & $\sim$260 [19] & Ca & [20]       \\
HD\,109573 & HR\,4796       & A0V   & $\sim$8 [20] & Ca   &  [20, 25]    \\
$\rho$~Vir & HD\,110411     & A0V  & $\sim$100 [20] &  Ca  &  [12]       \\
$\kappa$~Cen & HD\,132200   & B2IV   & 18.2$\pm$3.2 [26] & Ca  &  [27]      \\
HD\,145964  & HR\,6051      & B9V   & 9.2 [28] & Ca   &  [12]       \\
HD\,148283 & HR\,6123        & A5V   & 783$\pm^{229}_{412}$ [17] &  Fe  &  [29]    \\
HD\,156623 & HIP\,84881      & A0V   & $\sim$10 [30] & Ca   &  [27]      \\
5\,Vul & HD\,182919          & A0V    & 198 [31] & Ca  &  [8, 32]       \\
c\,Aql & HD\,183324          & A0IV   & 140 [31] & Ca  &  [33, 25, 32]   \\
\multirow{ 2}{*}{2\,And} & \multirow{ 2}{*}{HD\,217782}    & \multirow{ 2}{*}{A3V} & \multirow{ 2}{*}{100$\pm^{309}_{88}$ [17]} & Ca,  Fe, Cr, & \multirow{ 2}{*}{[34, 8, 32]}  \\
& & & & Mn, O & \\
HD\,24966         &  HIP\,18437 & A0V   & $\sim$10 [35] & Ca   &  [36]      \\
HD\,38056          & HR\,1966 & B9.5V  & $\sim$250 [37] & Ca  &  [36]      \\
$\theta$~Hya & HD\,79469    & B9.5V  & $\sim$90 [38] & Ca  &  [36]      \\
HD\,225200        & HR\,9102  & A1V    & $\sim$90 [39] & Ca  &  [36]      \\
$\alpha$~Pic$^\dagger$ & HR\,2550    & A7IV  & 519$\pm^{344}_{333}$ [17] &  Ca  &  [22]    \\
HD\,138629$^\dagger$ & HR\,5774       & A5V   & 24.8 [40] & Ca, Na   &  [41]       \\
\multirow{ 2}{*}{HD\,256$^\ddagger$} & \multirow{ 2}{*}{HR\,10}   & \multirow{ 2}{*}{A2IV/V}  & \multirow{ 2}{*}{$\sim$800 [23]} & \multirow{ 2}{*}{Ca, Fe, Na} &  [42, 18, 23, \\
& & & & & 43, 44] \\
$\phi$~Leo$^\ddagger$ & HD\,98058   & A5V & 500-900 [38] & Ca & [43, 45]     \\
\hline
\end{tabular*}
\end{center}
References: [1]\,\cite{MamajekBell2014}, [2]\,\cite{Ferlet_1987}, [3]\,\cite{Kiefer2014}, [4]\,\cite{Vidal-Madjar2017}, [5]\,\cite{Kiefer2023}, [6]\,\cite{Kiefer2014b}, [7]\,\cite{Grady2018}, [8]\,\cite{Montgomery2012}, [9]\,\cite{Miles2016}, [10]\,\cite{roberge2014}, [11]\,\cite{Roberge2008}, [12]\,\cite{Welsh2013}, [13]\,\cite{Kalas2005}, [14]\,\cite{Redfield2007a}, [15]\,\cite{Torres2006}, [16]\,\cite{Iglesias2019}, [17]\,\cite{Gullikson2016}, [18]\,\cite{Lecavelier1997b}, [19]\,\cite{Vican2012}, [20]\,\cite{Welsh2015}, [21]\,\cite{Montesinos2009}, [22]\,\cite{Hempel2003}, [23]\,\cite{Redfield2007b}, [24]\,\cite{Welsh1998}, [25]\,\cite{Iglesias18}, [26]\,\cite{Tetzlaff2011}, [27]\,\cite{Rebollido18}, [28]\,\cite{Gratton2023}, [29]\,\cite{Grady1996}, [30]\,\cite{Kral2017}, [31]\,\cite{Chen2014}, [32]\,\cite{Rebollido2020}, [33]\,\cite{Montgomery2017}, [34]\,\cite{Cheng2003}, [35]\,\cite{Rhee07}, [36]\,\cite{welsh2018}, [37]\,\cite{Morales2016}, [38]\,\cite{David_Hillenbrand2015}, [39]\,\cite{Su06}, [40]\,\cite{Fouesneau2022}, [41]\,\cite{Lagrange-Henri1990b}, [42]\,\cite{Lagrange-Henri1990}, [43]\,\cite{Eiroa16}, [44]\,\cite{Montesinos2019}, [45]\,\cite{Eiroa2021}, [46]\,\cite{BendahanWest2025}, [47]\,\cite{Deleuil1993}, [48] \cite{Vrignaud2024b}, [49] \cite{Vrignaud2025}, [50] \cite{Jolly1998}, [51] \cite{Roberge2000}.
\\
* Confirmed exocomet detection.\\
$^\dagger$ Unclear detection.\\
$^\ddagger$ False positive detection.
\\
\normalsize
\end{table*}

%% file: TablePhoto.tex
\begin{table*}
\begin{center}
\caption{\label{tab:exocomet_candidatesPH}Stars that show photometric signatures suggestive of an exocomet transit. This Table was based on Tables 1 and 2 in \cite{Strom2020}. For the stars with approximate spectral types, this is after applying reddening corrections with Gaia DR3. Stars missing stellar parameters have not been characterised yet.}

\begin{tabular*}{\textwidth}{@{\extracolsep\fill}llccr}
Name  & Other ID     & Sp. Type & Age [Myrs]  & Reference      \\
\hline
$\beta$~Pic* & HD\,39060    & A6V & 23$\pm$3 [1] & [2, 3, 13, 14]   \\
HD\,172555*   & HR\,7012 & A7V & 20$\pm$5 [11]  & [11]       \\
KIC\,3542116 & TYC\,3134-1024-1 &  F2V & 1725$\pm$100 [4] & [5, 12]       \\
KIC\,11084727 & BD+48\,2901 & F2V & 1918$\pm^{106}_{190}$ [4]  & [5, 12]      \\
KIC\,8462852 &  Boyajian's Star & F3V & $\sim$1000  [8]   & [6, 7, 8, 9, 10] \\
KIC\,8027456 & HD\,182952 & A1V & $\sim$100 [12] & [12] \\
TIC\,280832588 & 2MASS\,02591084-6624353 & $\sim$F5V & - & [14]\\
TIC\,73149665 & TYC\,8705-00361-1 & $\sim$F & - & [14]\\
TIC\,143152957 & 2MASS\,07473854-4234062 & $\sim$G4V & - & [14]\\
TIC\,110969638 & 2MASS\,07390295-2812217 & $\sim$G & - & [14]\\
TIC\,229790952 & UCAC4\,797-026470 & $\sim$K & - & [14]\\
RZ Psc & TYC 1753-1498-1 & K0V & 30 - 50 & [15] \\

\hline
\end{tabular*}
\end{center}
References: [1] \cite{MamajekBell2014}, [2] \cite{Zieba2019}, [3] \cite{Pavlenko2022}, [4] \cite{Queiroz2023}, [5] \cite{Rappaport2018}, [6] \cite{boyajian16}, [7] \cite{bodman16}, [8] \cite{Kiefer_2017}, [9] \cite{deeg18}, [10]\,\cite{wyatt2018}, [11] \cite{Kiefer2023}, [12] \cite{Kennedy_2019}, [13] \cite{Lecavelier22}, [14] \cite{Norazman2025}, [15] \cite{Gibson2025} \\
* Confirmed exocomet detection.\\
\end{table*}

%% file: TablePhotoWD.tex
\begin{sidewaystable}
\caption{\label{tab:exocomet_candidatesPHWD}White dwarfs showing photometric signatures that could be attributed to exocometary transits.}   
\begin{tabular*}{\textwidth}{@{\extracolsep\fill}llclllccr}
Name   & Sp. Type & Total Age   & Period & Pollution & Gas & Dust & Reference      \\
& & [Gyrs] & & & & & \\
\hline
ZTF J0328-129 & DBZ & 3.1 & 9.937 hr, 11.2 hr & Ca & Na$^{a}$ & N & [1]\\
WD J0923+7326 & DAZ & 1.1 & - & Mg, Ca & - & - & [2] \\
WD J1013-0427 & DBAZ & 2.7 & - & Mg, Si, Ca & Ca, O, Mg$^{e}$ & - & [2]\\
ZTF J1039+5245 & DAZ & 1.9 & 107.2 day & Ca & Ca(?)$^{a}$ & - & [3]\\
WD 1054-226 & DAZ & 4.6 & 25.02 hr & Mg, Al, Ca, Fe & N & N & [4,5]\\
\multirow{ 2}{*}{WD 1145+017} & \multirow{ 2}{*}{DBAZ} & \multirow{ 2}{*}{1.6} & \multirow{ 2}{*}{4.5-4.9 hr} & C, O, Mg, Al, Si, P, S, Ca,  & C, O, Na, Mg, Al, Ca,  & \multirow{ 2}{*}{Y} & \multirow{ 2}{*}{[6, 7]} \\
 & & & & Ti, V, Cr, Mn, Fe, Co, Ni, Cu & Ti, Cr, Mn, Fe, Ni$^{a}$  &  & \\
WD 1232+563 & DBAZ & 1.3 & - & O, Mg, Si, Ca, Ti, Cr, Mn, Fe & N & N & [8, 9]\\
WD J1237+5937 & DAZ & $\approx$ 10 & - & Mg, Al, Ca, Fe & - & - & [2] \\
WD J1302+1650 & DBAZ & 1.4 & - & Ca & - & - & [2]\\
WD J1650+1443 & DAZ & 1.7 & - & Mg, Ca & - & - & [2]\\
WD J1944+4557 & DA & 0.6 & - & - & Ca & - & [2]\\
\hline
\end{tabular*}
Total age includes both white dwarf cooling age and main sequence lifetime. It is calculated using \texttt{wdwarfdate} \citep{Kiman2022}.\\
$^a$ circumstellar absorption line \\
$^e$ circumstellar emission line \\
References: [1] \citet{Vanderbosch2021},
[2] \citet{Bhattacharjee2025}, 
[3] \citet{Vanderbosch2020}, 
[4] \citet{Farihi2022}, 
[5] \citet{VennesKawka2013},
[6] \citet{Vanderburg2015},
[7] \citet{LeBourdais2024}, 
[8] \citet{Hermes2024}, 
[9] \citet{Xu2019b}
\end{sidewaystable}